\documentstyle[12pt]{article}

\newcommand{\newsection}{    
\setcounter{equation}{0}
\section}
\def\appendix#1{
  \addtocounter{section}{1}
  \setcounter{equation}{0}
  \renewcommand{\thesection}{\Alph{section}}
  \section*{Appendix \thesection\protect\indent #1}
  \addcontentsline{toc}{section}{Appendix \thesection\ \ \ #1}
  }
\newcommand{\tr}[1]{\,{\rm tr}\,#1\,}
\def\e{{\,\rm e}\,}
\def\pint{\int\hspace{-1.07em}\not\hspace{0.6em}}

\def\eop{\vspace*{\fill}\pagebreak}
\def\be{\begin{equation}}
\def\ee{\end{equation}}
\def\bea{\begin{eqnarray}}
\def\eea{\end{eqnarray}}

\def\qq#1{\sqrt{#1^2+4c}}

\def\qqqq#1{\sqrt{#1^2+4b#1+4c}}

\def\L{\Lambda}
\def\l{\lambda}

\newcommand{\ie}{{\it i.e.}\ }

\newcommand{\eg}{{\it e.g.\ }}
\newcommand{\etal}{{\it et al\/.\ }}

\newcommand{\dd}[1]{{\partial \over \partial #1}}
\newcommand{\ddt}[1]{{\partial \over \partial t_{#1}}}

\newcommand{\half}{{\textstyle{1\over 2}}}
\renewcommand{\d}{{{\partial}}}
\newcommand{\p}{^{\prime}}

\newcommand{\ra}{\rightarrow}
\newcommand{\la}{\rightarrow}

\newcommand{\KK}{{\cal K}}

\newcommand{\PP}{{\cal P}}

\newcommand{\OO}{{\cal O}}

\newcommand{\fr}[2]{{\textstyle {#1 \over #2}}}

\title{{\bf \mbox{} \\A Hint on the External Field Problem for Matrix
Models}\vspace{.5cm}}
\author{{\bf L. Chekhov}\thanks{E--mail: \ chekhov@qft.mian.su}
\date{ }
\vspace{.5cm} \\
{\it Steklov Mathematical Institute} \\
{\it Vavilov st.42, GSP-1, 117966 Moscow, Russia}\\ \\
and \\ \\
{\bf Yu. Makeenko}\thanks{E--mail: \ makeenko@nbivax.nbi.dk \ \ / \ \
makeenko@desyvax.bitnet}
\vspace{.5cm} \\{\it The Niels Bohr Institute} \\
{\it Blegdamsvej 17, DK-2100 Copenhagen, Denmark} \\
{\it and} \\
{\it Institute of Theoretical and Experimental Physics} \\
{\it B.Cheremuskinskaya 25, 117259 Moscow, Russia}}

\begin{document}

\maketitle

\vspace{-16.6cm}

\begin{flushright}
hepth@xxx/9202006 \\ January 31, 1992
\end{flushright}

\vspace{13.8cm}

\begin{abstract}
We reexamine the external field problem for $N\times N$ hermitian one-matrix
models.
We prove an equivalence of the models with the potentials
$\tr{(\fr {1}{2N} X^2 + \log X - \L X)}$ and $\sum_{k=1}^\infty t_k
\tr{X^k}$ providing the matrix $\L$ is related to $\{t_k\}$ by
$t_k=\fr 1k \tr{\L^{-k}}-\fr N2 \delta_{k2}$.
Based on this equivalence we formulate a method for calculating the
partition function by solving the Schwinger--Dyson equations
order by order of genus expansion.
Explicit calculations of the partition function and of correlators
of conformal operators with the puncture operator are presented in genus one.
These results support the conjecture that our models are associated
with the $c=1$ case in the same sense as the Kontsevich model describes $c=0$.
\end{abstract}

\noindent
Submitted to {\sl Physics Letters B}

\eop

\newsection {Introduction}

The external field problem for matrix models has received recently much
attention due to the work by Kontsevich \cite{Kon91}. Generically, the
($N\times N$ hermitian) one-matrix model in an external field is defined by the
partition function
\be
Z[\L;N] = \int DX\, \e^{N\tr{(\L X-V_0(X))}}
\label{1}
\ee
where $V_0(X)$ is some potential. As follows from the results by
Kontsevich, this model with a cubic potential $V_0(X)\propto X^3$ describes
Witten's formulation \cite{Wit90} of 2D topological gravity.

The partition function (\ref{1}) can be calculated by the standard methods
of solving matrix models. While the orthogonal polynomial technique can
not be used due to the presence of the matrix $\L$, the method
\cite{BN81} of Schwinger--Dyson equations written in terms of
eigenvalues of $\L$ has been applied recently \cite{GN91,MS91} to
explicitly solve the Kontsevich model by genus expansion. An analogous
solution in genus zero has been obtained \cite{CM92} for the case of
the Kostov--Metha potential \cite{KM87}
\be
V_0(X)=\half X^2 - \alpha \log X ,
\label{2}
\ee
when it is associated with an external field problem for the Penner
model \cite{Pen86}.

The external field $\L$ in Eq.(\ref{1}) plays a role of a source for
correlators of $\tr{X^k}$ which can be obtained by differentiating
w.r.t.\ $\L$ and putting then $\L=0$. An alternative way to calculate
these correlators is to introduce the most general potential
\be
V(X)=\sum_{k=0}^\infty t_kX^k
\label{3}
\ee
and to consider the partition function
\be
Z[t_.;N] = \int DX\, \e^{-\tr{V(X)}}
\label{4}
\ee
whose derivatives w.r.t.\ $t_k$'s reproduce the same correlators of
$\tr{X^k}$'s provided one puts \linebreak
$V(X)=NV_0(X)$ after the differentiation.
In this approach, the role of an external field is played by the set of
couplings $\{t_k\}$. It is evident that the two ways of introducing an
external field are equivalent. As far as we know, this equivalence has
not been elaborated, however, in the literature.

In the present paper we prove the {\it exact\/} relation between the partition
functions (\ref{1}) and (\ref{4}):
\be
Z[\L;N]=\e^{\frac N2 \tr{\L^2}}Z[t_.;\alpha N],
\label{5}
\ee
which is valid provided $\L$ and $\{t_.\}$ are related by the
Miwa--type transformation
\be
t_k= \fr 1k \tr{\L^{-k}}-\fr N2 \delta_{k2} \hbox{\ \ \ \ for \ \ }n\geq1,
\,\,\,\,\, t_0=\hbox{\rm tr}\,\log{\Lambda^{-1}}
\label{6}
\ee
and $N\rightarrow \infty$ which makes $t_k$'s to be independent
variables. The proof is based on the fact that the Schwinger--Dyson
equations for both models are equivalent. Based on this equivalence,
we formulate then a method for
calculating the partition functions order by order of genus expansion
which is closed in spirit to that proposed by Gross and Newman
\cite{GN91} for the unitary matrix model and for the hermitian one with
a cubic potential. We calculate explicitly the partition function in
genus one and apply the result to calculate genus one contribution to
correlators of conformal operators with the puncture operator.
Our genus one results support
the conjecture \cite{CM92} that the model (\ref{1}) with the potential
(\ref{2}) (and therefore the model (\ref{4}) with the potential (\ref{3})\/)
is associated with the $c=1$ case, if $\alpha$ is identified with the
cosmological constant,
in the same sense as the Kontsevich model is associated with $c=0$.

\newsection{The Virasoro constraints}

A simplest way to prove the equivalence between the partition functions
(\ref{1}) and (\ref{4}) is to show that the Schwinger--Dyson equation
for the model (\ref{1}) with the potential (\ref{2}),
\be
\left\lbrace {\d^2 \over \d\L^2} +N\L \dd \L -\alpha N^2 \right\rbrace
\,\e^{-\frac N2 \tr{\L^2}}Z[\L]=0,
\label{7}
\ee
is equivalent to the set of the Virasoro constraints,
\be
L_nZ[t_.]= 0 \hbox{\ \ \ \ \ for\ \ \ }n\geq -1
\label{vc}
\ee
where
\be
L_n = \sum_{k=0}^\infty k t_k \ddt{n+k} + \sum_{k=0}^n \ddt{k}\ddt{n-k},
\label{vo}
\ee
imposed on the partition function (\ref{4}).

Eq.(\ref{7}) has been discussed in Ref.\cite{CM92}. The extra gaussian
factor in front of $Z$ is introduced to cancel one arising from the
integral (one can easily see this is the case of $\alpha=0$). The
commutation of this gaussian factor with the derivatives makes the sign
of the second term in parenthesises opposite to what would be naively
expected. As for the Virasoro constraints (\ref{vc}), (\ref{vo}), they
were obtained in Refs.\cite{AJM90} using the
invariance of the integral (\ref{4}) under the shift
$X\rightarrow X+\epsilon_n X^{n+1}$.

Now our purpose is to make a change of variables from (the eigenvalues
of) $\L$ to $\{t\}$ in Eq.(\ref{7}) similarly to what was done in
Refs.\cite{MMM91,GN91b} to derive the continuum Virasoro constraints
from the Kontsevich matrix model (in that case only odd $k$'s entered
the Kontsevich--Miwa transformation (\ref{6})). The only what we need is
the chain rule
\be
\dd{\L} = - \sum_{k=0}^\infty \L^{-k-1} \ddt{k}
\label{cr}
\ee
which is a consequence of Eq.(\ref{6}).

Making this change of variables, one obtains from Eq.(\ref{7})
\be
\sum_{n=-1}^\infty \L^{-n-2} L_n \e^{-\frac N2 \tr{\L^2}}Z=0
\ee
with $L_n$ given by Eq.(\ref{vo}) provided that
\be
\ddt{0}Z=-\alpha NZ
\label{norm}
\ee
which emerges formally for $n=-2$.
The last equation coincides with the normalization condition for $\alpha
N\times \alpha N$ hermitian one-matrix model. We have reproduces,
therefore, exactly the Virasoro constraints (\ref{vc}), providing the
partition functions are related by Eq.(\ref{5}).

The fact that the Virasoro constraints (\ref{vc}), (\ref{vo}) can be
obtained from the Schwinger--Dyson equation for the Kontsevich model
making the transformation (\ref{6}) with both even and odd $k$'s
was noticed by Marshakov, Mironov and Morozov \cite{MMM91}. It was not
recognized, however, that the proper external field problem is given by
the partition function (\ref{1}) with the potential (\ref{2}).

\newsection{Loop equations}

Loop equations are successfully applied for studying the matrix models
since the work by Kazakov \cite{Kaz89} (for a review of recent results,
see \eg Ref.\cite{Mak91}).

An insertion of the one-loop operator (with $\lambda$ being the Laplace
transformed momentum which corresponds to a loop of the length $l$) is given by
\be
{\delta \over \delta V(\lambda)}= - \sum_{k=0}^\infty \lambda^{-k-1} \ddt{k}.
\label{ddV}
\ee
This operator can be rewritten in terms of $\d / \d \L$.
Due to Eq.(\ref{6}), one gets
\be
\dd{\lambda_i}= - \sum_{k=0}^\infty \lambda_i^{-k-1}\ddt{k}
\label{ddli}
\ee
where $\lambda_i$ stands for an eigenvalue of $\L$.
As $N\rightarrow\infty$ when Eq.(\ref{ddli}) becomes exact, one can
define the density of eigenvalues of $\L$:
\be
\rho(\lambda)=\frac 1N \sum_{i=1}^N \delta(\lambda-\lambda_i)
\ee
so that
\be
{\delta \over \delta V(\lambda)}= \frac 1N
\dd{\lambda}{\delta\over\delta\rho(\lambda)}.
\label{3.4}
\ee

Similarly, one can relate (the derivative of) the loop source
$V(\lambda)$, which is defined by Eq.(\ref{3}), to $\rho$. Using
Eq.(\ref{6}), one gets
\be
V^\prime (\l)=\sum_{k=1}^\infty kt_k \l^{k-1}
=\sum_{j=1}^N \frac {1}{\l_j-\l} - N\l ,
\label{dhilbert}
\ee
where the sum over $k$ is convergent if $\l<\hbox{min}\mid
\l_i\mid$. As $N\ra\infty$ Eq.(\ref{dhilbert}) determines $V^\prime$
to be the Hilbert transform of $\rho$:
\be
\frac 1N V^\prime (\l)=
\pint dx {\rho(x)\over x-\l} -\l  .
\label{hilbert}
\ee
This equation expresses unambiguously each term of the $\l$-expansion of
$V^{\p}(\l)$
via $\rho$ provided the support of $\rho(x)$ vanishes at some finite
interval which includes the point $x=0$ (this is an analog of the above
condition $\l<\hbox{min}\mid \l_i\mid$). All the integrals below are
rigorously defined providing $\rho(\l)$ possesses this property.
The discussion of general properties of Eq.(\ref{hilbert}) is
beyond the scope of the present publication.

We compare now two equations. The first one is the Schwinger--Dyson
equation for the model (\ref{1}) written in terms of $\l_i$ \cite{CM92}
\be
\left\lbrace {\d^2 \over \d\l_i^2}+\sum_{j\neq i} \frac {1}{\l_i-\l_j}
\left(\dd{\l_i}-\dd{\l_j}\right)
+ N\l_i\dd{\l_i} -\alpha N^2 \right\rbrace \e^{\frac N2 \sum_k \l_k^2}
Z[\l_.] = 0.
\label{sd}
\ee
The second one is the loop equation written for the one-loop average
\be
W(\l)= {\delta \over \delta V(\lambda)} \log{Z[t_.]}
\label{W}
\ee
which reads (see, \eg \cite{Mak91})
\be
\int_{C_1} {d\omega\over 2\pi i}{V^\prime(\omega)\over (\l-\omega)}
W(\omega)=
(W(\l))^2 + {\delta \over \delta V(\lambda)}W(\l) .
\label{le}
\ee
This equation is equivalent to the Virasoro constraints (\ref{vc}),
(\ref{vo}) which can be obtained by expanding both sides of
Eq.(\ref{le}) in $1/\l$.

It follows immediately from Eqs.(\ref{ddV}), (\ref{3.4}), (\ref{W}) that
the r.h.s. of Eq.(\ref{le}) coincides with the first term (with the
double derivative) on the l.h.s.\ of Eq.(\ref{sd}).
In order to compare the remaining terms, one inserts Eq.(\ref{hilbert})
into the l.h.s.\ of Eq.(\ref{le})
and calculates the contour integral over $\omega$ by taking residuals at
the two poles. Applying again Eqs.(\ref{ddV}), (\ref{3.4}), (\ref{W}),
one gets exactly the remaining terms on the l.h.s. of Eq.(\ref{sd})
providing the normalization
condition (\ref{norm}) is satisfied. This completes the proof of the
equivalence between Eqs.(\ref{sd}) and (\ref{le}).

\newsection{Comparison of genus zero solutions}

It is instructive to compare the known (one-cut) genus zero solution of
loop equation (\ref{le}) with that of Eq.(\ref{sd})
which has been obtained in Ref.\cite{CM92}.

The genus zero solution to Eq.(\ref{le}), which was first obtained for a
general $V(\l)$ that involves both even and odd powers of $\l$ by Migdal
\cite{Mig83},  reads
\be
W_0(\l)=\int_{C_1}{d\omega\over 4\pi i}{V^\prime(\omega)\over (\l-\omega)}
\frac {\qqqq \l}{\qqqq \omega}
\label{W0}
\ee
with $b$ and $c$ given by
\be
\int_{C_1}{d\omega\over 2\pi i}{V^\prime(\omega)\over \qqqq \omega } =0,
\,\,\,\,\,\,
\int_{C_1}{d\omega\over 2\pi i}{\omega V^\prime(\omega)\over \qqqq \omega }
=2 \alpha N .
\label{bandc}
\ee
Inserting Eq.(\ref{hilbert}) and doing the contour integral as before,
one gets from Eq.(\ref{W0})
\be
\fr 1N W_0(\l)= \half \left[\qqqq{\l}-\l -\pint dx{\rho(x)\over x-\l}
\left( \frac{\qqqq \l}{\qqqq x}-1  \right)
\right]
\label{W0p}
\ee
while Eq.(\ref{bandc}) yields
\be
\half \int dx{\rho(x)\over \qqqq x}+ b=0\, ,\,\,\,\,
\half \int dx{x\rho(x)\over \qqqq x}+ c-3 b^2=\alpha +\half .
\label{bandcp}
\ee

One sees that (\ref{W0p}), (\ref{bandcp}) coincide with the
corresponding formulas of Ref.\cite{CM92}. The $W(x)$ which was
introduced there is related to our $W_0$ by
$W(x)=\frac 1N W_0(\l)+ \frac {\l}{2}$ while the definitions of $b$ and
$c$ are the same.
One more formula which would be useful for applications can be obtained
by differentiating Eq.(\ref{W0p}) (or (\ref{W0})\/) w.r.t.\ $\alpha$:
\be
\frac 1N {dW_0(\l)\over d\alpha}= {1\over\qqqq \l}.
\ee
Being expanded in $\frac {1}{\l}$, this expression reproduces Eq.(5.20)
of Ref.\cite{CM92}. An advantage of the approach of Ref.\cite{CM92} is
that it allowed to calculate $\log{Z}$ itself.

For an iterative solution of Eq.(\ref{sd}) which is considered below, we
shall need an explicit expression for the irreducible two-loop
correlator
\be
W(\l,\mu)={\delta \over \delta V(\mu)} {\delta \over \delta V(\lambda)}
\log Z .
\ee
Applying Eq.(\ref{3.4}) to $W_0$ given by Eq.(\ref{W0p}), one gets in
genus zero
\be
W_0(\l,\mu)= \frac{1}{2(\l-\mu)^2}\frac {\l\mu-2b(\l+\mu)+4c}
{\qqqq \l \,\,\qqqq \mu},\,\,\,\,W_0(\l,\l)=\frac{b^2-c}{(\l^2+4b\l+4c)^2}
\label{W20}
\ee
which coincides with the known result by Ambj{\o}rn \etal \cite{AJM90}.

The solution (\ref{W0p}), (\ref{bandcp}) is simplified if $\rho(\l)$ is
a symmetric function $\rho(\l)=\rho(-\l)$. The first equation in
(\ref{bandcp}) yields then $b=0$ while only even powers of $x$ enter
the second one (as well as Eq.(\ref{W0p})\/). This case corresponds to
the so-called reduced hermitian one-matrix model, \ie to vanishing odd
times $t_{2m+1}$. It implies for the Miwa transformation (\ref{6})
that only even $k$'s are present. This situation is complementary to the
original Kontsevich model when only odd $k$'s appear. The loop
equations for the reduced hermitian one-matrix model can not be
formulated, however, entirely in terms of the even times. The odd times
inevitably appear for this model to higher orders. This does not happen
for the case of the {\it complex\/} matrix model (see, \eg \cite{Mak91})
which is formulated entirely via the even times. It is the model which
is complementary in this sense to the Kontsevich model. A study of the
external field problem for the complex matrix model will be published
elsewhere.

\newsection{The genus one solution}

Our idea of how to solve Eq.(\ref{sd}) (or (\ref{le})\/) iteratively is
similar to one which has been proposed by Gross and Newman \cite{GN91}
for the unitary matrix model and for the hermitian one with
a cubic potential.

Let us introduce the new variables
\bea
B_p&=& 2^{p-1}(2p-1)!! \int d\l \rho(\l){1\over
\left(\l^2+4b\l+4c\right)^{p+\half}}, \nonumber \\
C_p&=&  2^{p-1}(2p-1)!! \int d\l \rho(\l)
{\l\over \left(\l^2+4b\l+4c\right)^{p+\half}}
-(\alpha + \half) \delta_{p0}
\label{nv}
\eea
where $b$ and $c$ are determined by Eq.(\ref{bandcp}) which reads now
\be
 B_0+b=0, \,\,\,\, C_0+c- 3 b^2= 0
\label{bandcn}
\ee
(\/$(-1)!!=1$ by definition). Let us define $F_g$ ---
genus $g$ contribution to $\log Z$ --- by the formula
\be
\log Z = \sum_{g=0}^\infty N^{2-2g} F_g.
\ee
Now our conjecture is that $F_g$ would depend
at $1\leq g <\infty$ only on $B_p$ and $C_p$ for $p\leq P$ where $P$ is
some finite number (the larger $g$ the larger $P$). This is in contrast
to the $t$-dependence of $F_g$ which always depends on the whole set
$\{t_k\}$. Such a behavior of $F_g$ for the  Kontsevich model has been
advocated recently by Itzykson and Zuber \cite{IZ92}.

Let us find explicitly the genus one correction to the genus zero
results which are described in the previous section.
Defining
\be
W_1(\l) = {\delta \over \delta V(\l)} F_1
\ee
with $\delta / \delta V(\l)$ given by Eq.(\ref{3.4})
and substituting into Eq.(\ref{sd}), one gets the following linear
equation for $W_1(\l)$:
\be
(\fr 2N W_0(\l)+\l)W_1(\l)+ \pint dx \rho(x)
\frac {W_1(\l)-W_1(x)}{\l-x} + \fr 1N W_0(\l,\l) =0,
\label{sd1}
\ee
where $W_0(\l)$ and $W_0(\l,\l)$ are given by Eqs.(\ref{W0p}) and
(\ref{W20}), respectively.

For an arbitrary $\rho$, the solution of Eq.(\ref{sd1}) requires tedious
calculations. Some simplifications occur if $\rho$ is a symmetric
function $\rho(\l)=\rho(-\l)$. As is discussed above, this
corresponds to the reduced hermitian matrix model (\ie to an even
potential $V$). For this case, one gets $b=0$ and $B_p=0$ in
Eqs.(\ref{nv}) and (\ref{bandcn}) which simplifies calculations.
Using the property $\rho(\l)=\rho(-\l)$, Eq.(\ref{W0p}) can be rewritten
as
\be
\fr 2N W_0(\l)+ \l = \qq \l - \pint dx\rho(x) \frac {x}{x^2-\l^2}
\frac {\qq \l}{\qq x}.
\label{W0even}
\ee
By a direct differentiation of this formula one gets
\be
\dd c (\fr 2N W_0(\l)+ \l)=\frac {2(1- C_1)}{\qq \l}
\label{dWdc}
\ee
which will be extensively used below. We shall need as well
the following rules of differentiation:
\bea
\frac {\d C_p}{\d c}= - C_{p+1},\,\,\,
\frac {\delta c}{\delta V(\l)}= {2c\over (\l^2+4c)^{\frac 32}(C_1-1)},
\nonumber \\
\frac {\delta C_1}{\delta V(\l)}=
\left[ \frac{cC_2}{1-C_1} -1 \right]\frac {2}{(\l^2+4c)^{\frac 32}}
+ {12c\over (\l^2+4c)^{\frac 52}},
\label{rules}
\eea
which are easy to derive from Eqs.(\ref{nv}), (\ref{bandcn}).

Using the property $\rho(\l)=\rho(-\l)$,
let us represent the second term on the l.h.s.\ of Eq.(\ref{sd1})
as the linear operator $\KK$:
\be
\KK W_1(\l)=\int dx \rho(x) K(\l,x)W_1(x)\equiv
\pint dx \rho(x) \frac{\l W_1(\l)-x W_1(x)}{\l^2-x^2}
\label{K}
\ee
In matrix notations, when $\l$ and $x$ are replaced by the eigenvalues
$\l_i$ and $\l_j$, this operator becomes an $N\times N$ matrix $\KK _{kl}$.
Such an operator was considered in Ref.\cite{GN91}.
Eq.(\ref{sd1}) can now be rewritten as
\be
(\fr 2N W_0(\l)+\l)W_1(\l) +\int dx \rho(x) K(\l,x)W_1(x)
= \fr 1N \frac{c}{(\l^2+4c)^2} .
\label{sd1p}
\ee
It is the equation which is solved below.

The form of the operator $\KK$ and the fact that $W_1(\l)=-W_1(-\l)$ suggest
the following ansatz
\be
W_1(\l)= \sum_{n=0}^\infty\frac{A_n}{(\l^2+4c)^{n+\half}}
\label{ans}
\ee
where the coefficients $\{A_n\}$ are functions of $\{C_p\}$.
It is easy to calculate of how the operator $\KK$ acts on the `basis
vectors' $1/(\l^2+4c)^{n+\half}$. One first calculates for $n=0$
and then obtains a general formula by applying $(\d / \d c)^n$. The
result reads
\be
\int dx \rho(x) K(\l,x)\frac{1}{(x^2+4c)^{n+\half}}=
\frac {(-1)^n}{2^n(2n-1)!!}\left(\dd c\right)^n
\left[ 1-\frac{2W_0(\l)+\l}{\qq \l}\right].
\ee
The term arising from the action of $(\d/\d c)^n$ on  $1/\qq \l$ equals to the
first term on the l.h.s.\ of Eq.(\ref{sd1p}) with the minus sign and
cancels it when inserted into Eq.(\ref{sd1p}). The other terms can be
easily calculated using Eq.(\ref{dWdc}). Calculating the derivatives,
one gets finally
\bea
& &\left( 2W_0(\l)+\l+\KK\right)\frac{1}{(\l^2+4c)^{n+\half}}=
\frac{1-C_1}{(\l^2+4c)^n}-
\sum _{p=2}^n \frac{C_p}{2^{p-1}(2p-1)!!}\frac{1}{(\l^2+4c)^{n-p+1}},
 \nonumber \\
& &\left( 2W_0(\l)+\l+\KK\right)\frac{1}{\qq \l}=1\hbox{\ \ \ \ \ \
\ \ \ \ \ \ \ for \ }
n=0
\eea
where the operator notation for the l.h.s.\ of Eq.(\ref{sd1p}) has been
used. This formula is similar to that by Gross and Newman \cite{GN91}
while the definition of the `moments' $C_p$ is different for our model.

One sees now that the r.h.s.\ of Eq.(\ref{sd1p}) is reproduced by the
$n=2$ term so that Eq.(\ref{sd1p}) is satisfied if all $A_n=0$
except
\be
A_1=\fr 1N \frac{C_2c}{6(1-C_1)^2},\,\,\,\,A_2=\fr 1N \frac{c}{1-C_1}.
\ee
Therefore, we have found the genus one solution to be
\be
W_1(\l)=\fr 1N \left\lbrace
\frac{C_2c}{6(1-C_1)^2}\frac{1}{(\l^2+4c)^{\frac 32}}
+\frac{c}{1-C_1}\frac{1}{(\l^2+4c)^{\frac 52}}\right\rbrace.
\label{W1}
\ee
It is worth noticing that we have obtained the explicit genus one solution
to the reduced model with an {\it arbitrary\/} potential $V(\l)=V(-\l)$.
Previously it was calculated \cite{AM90} by solving loop equations only
for the case of a polynomial potential with the highest power $\l^6$.
We have verified that the result for the quartic potential can be reproduced
by our formula (\ref{W1}) when $\rho(\l)$ is such that only $t_2$ and $t_4$
are nonvanishing so that
\be
C_1=1+2t_2-24t_4c,\,\,\,C_2=24t_4,\,\,\,C_p=0 \hbox{\ \ \ \ \ \ for \ }
p\geq3
\ee
and Eq.(\ref{bandcn}) for $c$ reads
\be
-2t_2c+12t_4c^2=\alpha.
\ee
The general formulas which express the `moments' $C_p$ via $\{t_{2k}\}$
for a polynomial $V(\l)$ look similar.

Using the collection of formulas from Eq.(\ref{rules}), one integrates
Eq.(\ref{W1}) to obtain
\be
F_1 = - \frac{1}{12}\log{\{ c\,(1-C_1)\}}.
\label{F1}
\ee
While our genus one result looks similar to that of
Ref.\cite{GN91,IZ92}, the coefficient in front of the logarithm is now
$\frac{1}{12}$ instead of $\frac{1}{24}$. As is discussed in the next
section this is related to the fact that our matrix model is associated
with $c=1$ and not with $c=0$.

\newsection{The relation to c=1}

We discuss in this section of how our genus one results support the
conjecture of Ref.\cite{CM92} (which was based on the genus zero
calculations)
that the model (\ref{1}) with the potential (\ref{2}) is associated
with $c=1$ providing $\alpha$ is identified with the cosmological
constant. This model has been called the Kontsevich--Penner model.

Let us start with the case of $\L\la\infty$ when the Kontsevich--Penner
model is reduced to the standard Penner model \cite{Pen86} which
corresponds \cite{CM92} to $b=0$, $c=\alpha$, $B_p=0$ and $C_p=0$ for
$p\geq 1$ in our notations. This case can be easily recovered by the
genus one solution from the previous section. One gets
from Eq.(\ref{F1})
\be
F_1=-\frac{1}{12} \log \alpha
\ee
which exactly reproduces the corresponding result of Distler and Vafa
\cite{DV91} for the Penner model. Notice that we did not take the
`double scaling limit' to obtain this result since our model is
associated with the {\it continuum\/} case.

The correlators of conformal operators $\OO_n$ with the puncture operator
$\PP$ can be obtained for our model
in genus one by applying $d/d\alpha$ to Eq.(\ref{W1}) and
expanding in $1/\l$. The only $C_0$ depends explicitly on $\alpha$ so
that our collection of formulas (\ref{rules}) should be supplemented by
\be
\frac {d c}{d \alpha}=\frac{1}{1-C_1}.
\label{dcda}
\ee
By a direct differentiation of Eq.(\ref{F1}) one gets
\be
\frac{d^2}{d \alpha^2} F_1 =\frac{1}{12(1-C_1)^2}
\left\lbrace \frac{1}{c^2}+ \frac{C_2}{c(1-C_1)}+ \frac{C_3}{(1-C_1)}
+ \frac{2C_2^2}{(1-C_1)^2}\right\rbrace
\ee
which corresponds to genus one contribution to the correlator of the two
puncture operators $\PP$.
One can combine this result with the genus zero calculation of
Ref.\cite{CM92} to give
\be
\frac{d^2}{d \alpha^2} F = \log{(-C)}
\label{susc}
\ee
where $C$ is defined by
\be
C=c+\frac{1}{12N^2(1-C_1)^2}
\left\lbrace \frac{1}{c}+ \frac{C_2}{(1-C_1)}+ \frac{cC_3}{(1-C_1)}
+ \frac{2cC_2^2}{(1-C_1)^2}\right\rbrace
\label{C}
\ee
The meaning of this formula is disscussed below.

To calculate the next correlators, one differentiates Eq.(\ref{W1})
w.r.t.\ $\alpha$. The result reads
\bea
N\frac{d}{d\alpha} W_1(\l)&=&
\left\lbrace \frac{C_2-cC_3}{(1-C_1)^3}- \frac{2cC_2^2}{(1-C_1)^4}
\right\rbrace  \frac{1}{6(\l^2+4c)^{\frac 32}} \nonumber \\
&+&\left\lbrace \frac{1}{(1-C_1)^2}- \frac{2cC_2}{(1-C_1)^3}
\right\rbrace\frac{1}{(\l^2+4c)^{\frac 52}}
-\frac{10c}{(1-C_1)^2}\frac{1}{(\l^2+4c)^{\frac 72}}.
\label{dW1da}
\eea
The explicit expressions for the correlators of $\OO_{2k}$ with $\PP$,
which can be obtained by expanding the r.h.s.\ of Eq.(\ref{dW1da}) in
$1/\l^2$, are rather involved while
some simplifications occur after the shift (\ref{C}). As an example we
present the sum of the genus zero and genus one results for the correlator
of $\PP$ and $\OO_2$ (the dilaton operator in the case of the Penner model)
which reads
\be
\langle \OO_2 \PP \rangle \equiv \frac {\d^2}{\d t_2 \d \alpha}F=
-2C + \frac {1}{6N^2(1-C_1)^2}
\left[ \frac{2C_2}{(1-C_1)}+ \frac 1c\right].
\ee
When expressed in terms of $\{t_{2k}\}$, this formula is to be compared
with the corresponding correlator for $c=1$ CFT.

Let us speculate now on an nonperturbative essence of our results.
Eq.(\ref{susc}) looks like the equation $\d^2 F/ \d t_0^2=u$ which
expresses the susceptibility $u$ via $F$ for the Kontsevich model while
Eq.(\ref{C}) is an analog of the string equation (to the given order of
the genus expansion) which relates $u$ to the `times' $\{t_k\}$. The
transformation (\ref{6}) looks like the transformation
which has been introduced  by Kharchev \etal
\cite{KMMMZ91} for the generalized Kontsevich
models to relate their partition functions to the (reduced) KP
$\tau$-function while differs from it by the fact that the role of the
cosmological constant is played now not by $t_0$ but by the `conjugate' (in
the sense of Eq.(\ref{norm})\/) variable $\alpha$. As is shown in
Ref.\cite{KMMMZ91}, the whole description of $c<1$ CFT's can be obtained in
this framework and the set of Virasoro and W-- constraints can be
constructed. Our conjecture indicates, therefore, that the set of the
Virasoro constraints (\ref{vc}), (\ref{vo}) which is constructed with the
use of the variable $t_0$ --- `conjugate' to the cosmological constant
$\alpha$ ---  can be used alternatively to describe a $c=1$ CFT.

One more argument in favour of our conjecture is a similarity between the
Virasoro constrains (\ref{vc}), (\ref{vo}) and those for the case of the
Witten's generalization \cite{Wit91b} of the intersection theory on moduli
space analytically continued to $k=-3$ (or to $p=-1$ in the notations of
Ref.\cite{Dij91}) when it is equivalent in a certain sense to the Penner
model.
It would be very interesting to study the relation between all these
approaches.


\end{document}